%% file: main.tex
\documentclass[preprint,12pt]{elsarticle}
\usepackage[T1]{fontenc}

\usepackage{graphicx}
\graphicspath{{./}}
\usepackage{amssymb}
\usepackage{booktabs}
\usepackage{caption}
\usepackage[labelformat=simple]{subcaption}
\usepackage{tikz}
\usetikzlibrary{shapes, arrows, positioning}
\usepackage{pgfplots}
\usepackage{pdfpages}
\usepackage{placeins}
\usepackage{xspace}
\usepackage{amsmath}
\usepackage{lipsum}
\usepackage{hyperref}
\usetikzlibrary{shapes.geometric, arrows}
\journal{Medical Engineering \& Physics}

\usepackage{dsfont}
\newcommand*{\IR}{\ensuremath{\mathds R}}

\renewcommand*{\vec}{\ensuremath{\boldsymbol}}
\newcommand*{\dx}{\ensuremath{\; \textup dV}}

\pgfplotsset{compat=1.18} 
\begin{document} 

\begin{frontmatter}



\title{Denture reinforcement via topology optimization}

%\def\correspondingauthor{\footnote{Corresponding author: rabia.altunay@lut.fi}}
\author[LUT,TUAS]{Rabia Altunay}
\ead{rabia.altunay@lut.fi}
\author[TUAS]{Kalevi Vesterinen}
\author[TUAS]{Pasi Alander}
\author[TUAS]{Eero Immonen}
\author[UDS,LUT]{Andreas Rupp}
\author[LUT]{Lassi Roininen}

% \author{Rabia Altunay$^{a,b}$, Kalevi Vesterinen$^{b}$, Pasi Alander$^{b}$, Eero Immonen$^{b}$, Andreas Rupp$^{a}$, Lassi Roininen$^{a}$}

 %\author{Rabia Altunay\corref{cor1}\fnref{label2}}
% \ead{rabia.altunay@lut.fi}
%\fntext[label2]{}
%\cortext[cor1]{}
\affiliation[LUT]{
    organization={School of Engineering Sciences, Lappeenranta-Lahti University of Technology LUT},
    addressline={Yliopistonkatu 34}, 
    %postcode={FI-53850},
    city={FI-53850 Lappeenranta},
    country={Finland}}

\affiliation[TUAS]{
    organization={Computational Engineering and Analysis Research Group, Turku University of Applied Sciences},
    addressline={Joukahaisenkatu 3}, 
    %postcode={FI-20520},
    city={FI-20520 Turku},
    country={Finland}}

\affiliation[UDS]{
    organization={Department of Mathematics, Saarland University},
    % addressline={Joukahaisenkatu 3}, 
    %postcode={FI-20520},
    city={DE-66123 Saarbr\"ucken},
    country={Germany}}


\begin{abstract}
 We present a computational design method that optimizes the reinforcement of dentures and increases the stiffness of dentures. Our approach optimally places reinforcement in the denture, which modern multi-material three-dimensional printers could implement. The study focuses on reducing denture displacement by identifying regions that require reinforcement (E-glass material) with the help of topology optimization. Our method is applied to a three-dimensional complete lower jaw denture. We compare the displacement results of a non-reinforced denture and a reinforced denture that has two materials. The comparison results indicate that there is a decrease in the displacement in the reinforced denture. Considering node-based displacement distribution, the reinforcement reduces the displacement magnitudes in the reinforced denture compared to the non-reinforced denture. The study guides dental technicians on where to automatically place reinforcement in the fabrication process, helping them save time and reduce material usage. 

 %The study aims to provide denture patients' comfort and quality of life.

%(12) (asl) [enhance denture patients' comfort and quality of life...]

%(14) (asl) To do this ----->  To this end

% (15)(asl) The comparison results indicate that the maximum displacement is a decrease in the reinforced denture. ----> The comparison results indicate that there is a decrease in the displacement in the reinforced denture.

% (asl) Moreover, node-based displacement distribution demonstrates that the average displacement distribution is substantially improved in the reinforced denture. ----> Is node-based displacement distribution sort of a method?

\end{abstract}
%


 %\item We present a computational design method for optimizing denture durability.
% \item This method addresses the reinforcement of weak areas in dentures.
 %\item The method also allows for optimization problems involving two distinct materials.
 %\item We reduce displacement via region-specific reinforcement.
% \item Our method improves denture strength, durability, and performance.


\end{frontmatter}

\section*{Abbreviations}
We abbreviate finite element method by FEM, polymethylmethacrylate by PMMA, solid isotropic material with penalization by SIMP, partial differential equation by PDE, computer-aided design by CAD, and computer-aided manufacturing by CAM.

\section{Introduction}
% 
%A fixed partial denture is a dental restoration to replace missing teeth by creating a permanent attachment to adjacent teeth or dental implants. Fixed partial dentures are similar to a bridge fixed onto abutment support structures. Clinical studies highlight the need to improve fixed partial dentures constructed by fiber-reinforced composites \cite{ShinyaY10,ShiF09,IranmaneshAGK14,ShiFQ08}. These enhancements are required to increase the durability of fixed partial dentures. That is, they address potential failures like the detachment of veneer material from the fiber reinforcement substructure, separation of the composite and tooth interface, or breakage of the pontic.

Structural design strategies exist to optimize the performance and functionality of biomedical devices by offering personalized customization \cite{imeko1}. During the design process, challenges include the bio-compatibility of the materials, safety regulations, long-term durability, and cost of the biomedical devices \cite{lecturenote1, compositereview}. Design principles such as generative design and BioTRIZ can be employed to manufacture customized biomedical devices and address the difficulties \cite{imeko2, BioTriz}. The engineering design principles are virtually always complemented by numerical methods, such as topology optimization and finite element method (FEM) \cite{toporeview,structuraldesign,sridhar2010optimal, chethan2019finite}. 

Topology optimization algorithms are computational methods used to determine the optimal material distribution of an object. They aim to optimize an objective function, such as stiffness, while considering constraints like mass or volume of the object \cite{topologyandshape}. The topology optimization of biomedical devices might require considering the interaction between the implant and tissue for bio-compatibility reasons. This interaction can be difficult to formalize in terms of optimization constraints. For example, topology optimization of dental implants has been studied to maximize their stiffness and bone ingrowth \cite{lecturenotes2}. Besides the design principles and numerical tools, additive manufacturing is another key component of biomedical device development, enabling the creation of personalized implants and prostheses tailored to individual patient anatomies \cite{sun2021additive, additive}. Improving the prediction and control of surface roughness in three-dimensional-printed devices is paramount to ensure their high quality. Quality control of the printed devices has been studied by examining the relationship between the printing parameters and surface smoothness \cite{lecturenotes3}.

Prostheses can be reinforced with cobalt-chromium alloy metal wires, aramid fiber, polyethylene fiber, carbon fiber, and glass fiber \cite{materials}. Currently, denture reinforcement is often made of E-glass. Fiber reinforcements increase the strength of removable acrylic dentures and prevent the breaking of prosthesis \cite{NarvaVHY01}. There are two ways to strengthen acrylic dentures: support the entire denture base plate with fiber reinforcement (total reinforcement) \cite{LadizeskyCW90} or place fiber reinforcement only in the weakest part of the denture (partial reinforcement) \cite{VallittuL92reinforcement}. The first approach is mainly made with a fiber reinforcement net (bidirectional fiber sheet), and it is generally used for upper dentures, where the denture base plate is as big as the palate. Partial fiber reinforcement is widely preferred and clinically proven for upper and lower dentures. The effective use of partial fiber reinforcement requires reliable information about existing or presumed paths of fracture lines \cite{NarvaVHY01}. 

%Unidirectionally oriented fiber is tailored to fit inside the denture. In our study, we use E-glass as fiber reinforcement. To perform numerical simulations, we must specify Young's modulus and Poisson's ratio of the materials. 

%Moreover, as unidirectional dental fiber reinforcements are frequently small in diameter, the reinforcement must be positioned correctly.

%A removable denture fracture is a typical fatigue failure of the brittle polymethylmethacrylate (PMMA) resin base, induced by the cumulative stress from hundreds of thousands of biting cycles annually \cite{fracture}. Such fractures account for approximately 64\% denture repairs involving damage to the acrylic resin base \cite{perea20213d}. Incorporating fiber reinforcements into removable acrylic dentures enhances their mechanical strength, thereby mitigating the fracture risk during clinical application \cite{perea20213d,altarazi2022assessing}.

%Different materials are used in dental prostheses, such as ceramics, metals, and, most commonly, acrylic resin \cite{materials}. 

Material properties are a critical factor in the performance of dentures. A removable denture fracture is a typical polymethylmethacrylate (PMMA) resin base failure induced by the stress of biting cycles \cite{fracture,reinforcement}. These fractures constitute approximately 64\% of denture repairs involving acrylic resin base damage \cite{perea20213d}. Several approaches have been introduced to increase the strength of dentures made of PMMA \cite{Zafar20} and decrease the stress distribution in the prosthetic structure \cite{JomjunyongRRACK17}. The results indicate that the type of fiber reinforcement
and thickness affect the resistance of the denture structure \cite{reinforcement1}. PMMA is commonly used clinically in dentures due to its bio-compatibility and ease of fabrication. It is well-tolerated by the human body and durable in the moist oral environment. It can be quickly processed into various shapes and sizes. Additionally, PMMA offers aesthetic benefits and has a long history of use in dental research and clinical practice \cite{Zafar20,perea20213d}.

In dentistry, FEM is used to model dental implants and to analyze fiber reinforcement of fixed partial dentures \cite{ShinyaY10} and bridges \cite{ShiF09}. Considerable research is undertaken to optimize the conventional fiber reinforcement designs. For example, the cavity shape can be optimized to minimize inter-facial stresses between the restoration and the tooth, thus reducing the likelihood of debonding \cite{ShiFQ08}. Two optimization methods are used: Stress-induced material transformation and stress-induced volume transformation \cite{ShiFQ08}. Also, \cite{ShiF09} optimizes the fiber reinforcement of a three-unit fixed partial denture to reduce stresses that can cause failure. The work uses structural optimization methods and FEM to gradually reinforce areas of fixed partial dentures with fiber reinforcement where high tensile stresses are identified via stress-induced material transformation. The maximum principal stress direction determines the fiber reinforcement orientation. Fiber reinforcement optimization studies using FEM are typically done on a case-by-case basis \cite{ShiF09, ShiFQ08}. These approaches vary the reinforcement's position, length, thickness, or shape systematically \cite{ShinyaY10}, which requires manual parametrization of the reinforcement designs. Such manual designing and evaluation of reinforcement candidates limits their clinical applicability.

%FIXME: Add porous lattice. %Triply periodic minimal surfaces offer a method to balance mechanical strength with permeability for implants requiring integration with biological tissues. Repetitive unit cells in design allow for creating complex structures that meet specific mechanical requirements \cite{compositereview, structuraldesign}. However, further investigation will elucidate the potential of bio-compatible multi-materials, ensuring regulatory compliance and economic competitiveness.

%further research is needed to develop bio-compatible multi-materials, ensure regulatory compliance, and reduce costs.

%For this reason, we propose a novel holistic topology optimization method that has a numerical optimizer scheme called solid isotropic material with penalization (SIMP) to optimize the dentures that places E-glass reinforcement such that the longevity of the denture is increased while, at the same time, keeping material costs for the reinforcements low. 

We propose a topology optimization method based on the solid isotropic material with penalization (SIMP) for reinforcements of three-dimensional dentures \cite{originalSIMP}. We limit the scope of the study to design the reinforcement computationally. The rationale of our approach is similar to a paper \cite{extension}, which considers the topology optimization of dental implants. That paper studied the optimization of dental implants made of one material. We extend that study to optimize the complete denture made of two materials. Our approach is based on the idea of our previous publication \cite{ECMS}. The approach eliminates the need to modify existing topology optimization routines. Using SIMP, we distribute the stronger material (reinforcement, E-glass) and assign lower-modulus material (resin base, PMMA) to non-reinforced areas within the denture. This approach is numerically robust, unlike other multi-material topology optimization methods \cite{hvejsel2011material}.  %This avoids the complexity of implementing separate multi-material optimization algorithms. 
The method enables the placement of E-glass reinforcement in dentures to increase their stiffness while keeping the material costs low for the reinforcement. 
%Implementing advanced tools like reinforced dentures with multi-material three-dimensional printers is promising and feasible in practical applications. 
%This study primarily focuses on a computational design approach for dental prostheses.
%It comes up with a new reinforcement optimization method in dentures, with limited emphasis on the feasibility of manufacturing or the physical production of the prostheses. 

%We complement the numerical results with a mesh sensiticyty study
%Challenges encountered in the study relate to the complexity of the model, which involves two different materials: PMMA  and E-glass. To address this issue, an approach is devised in which the optimization process begins with one material,and the results are validated with a mesh sensitivity study.

%The proposed method is more functional than traditional optimization methods \cite{ShinyaY10, ShiF09, ShiFQ08}. The method is designed to meet the required performance specifications while finding optimal reinforced regions in the denture by fixing the material used. 
Objectives of the study are as follows:

\begin{itemize}
    %\item The use of density-based topology optimization with FEM to remove excess material from the reinforcement to find optimal positions of it in the denture.
    \item We propose a non-parametric optimization approach to design reinforced dentures (multi-material).
    \item We decrease the compliance of the denture through the proposed approach. 
   
   % \item We assess the effectiveness of the reinforcement in reducing the displacement of the denture. 
   % \item Topology optimization of the reinforcement for dentures through subtracting specific mass is not previously investigated in published research studies.
   %\item We minimize compliance and reduce denture displacement by optimizing reinforcement topologically.
  % \item We perform a mesh sensitivity study of the proposed reinforcement optimization by validating the displacement of the prosthesis with several mesh sizes.
\end{itemize}

The rest of the paper is organized as follows. In Section 2, we propose the optimization method for reinforcement applications and specify the material properties for this study. In Section 3, we apply the proposed method in a numerical example of a three-dimensional denture and compare the reinforced and the non-reinforced dentures. We also perform a mesh convergence study to verify the accuracy of the numerical results. In Section 4, we discuss the limitations of our study and clinical perspectives. In Section 5, we conclude the general aspects of our study and suggest future research ideas.

% position ----> (optimal) placement
% (113) reduce denture displacement

%The novelty of this study proposes a computational method that is different from researchers who traditionally have done case-by-case optimization in the dental area to optimize reinforcement. We use a numerical optimizer scheme called solid isotropic material with penalization (SIMP) to optimize reinforcement hence the proposed method is a more holistic approach than traditional optimization methods. 

\section{Methodology}
% 
%We model the denture's reinforcement (strong material) and subtract the material to see the distribution of mass in the optimization region. Then, we fill non-optimized regions with soft material. In the end, we have two materials in the optimized model. Our mathematical model is based on the partial differential equations (PDE) for linear elasticity, frequently used to describe small displacements of elastic materials; see \cite{Slaughter12, Gurtin73} for further details.
%Partial differential equations (PDE) might have exact \cite{rabia21,rabia22,rabia23} or numerical solutions \cite{Qian93, Chen23, Zheng23}. 

%-displacement of the denture, Linear elasticity PDE, maybe the equation itself and compliance formula
%-Elliptic PDE, analytical solution for complex solution
%-FEM approximation of the PDE, linear basis functions 
%-Numerical tools Ansys
%-SIMP method  for minimization of the compliance

\label{section2}
We study the displacement $\vec u = \vec u(x,y,z)\colon \Omega \to \IR^3$ of the denture with and without reinforcement. Here, $\Omega\subset\IR^3$ denotes the three-dimensional domain of the denture. We assume that the prosthesis materials follow linear elasticity \cite{Gurtin73}. For linear materials, the displacement $\vec u(x,y,z)$ is governed by the linear elasticity partial differential equation (PDE) \cite{Slaughter12}, which is
\begin{subequations}
	\begin{align}
		\label{pde}
		- \nabla \cdot (\vec C : \vec \epsilon(\vec u ) ) & = \vec 0 && \text{ in } \Omega,\\
		\vec u & = \vec 0 && \text{ on } \Gamma_D,\\
		\vec C : \vec \epsilon(\vec u) \vec n & = \vec f && \text{ on } \Gamma_N, %\\
		% C \epsilon(\vec u) \cdot  \vec n & = \vec 0 && \text{ on } \partial \Omega \setminus \Omega_D \cup \Omega_N
	\end{align}
\end{subequations}
where the strain tensor $\vec \epsilon$ is $\vec \epsilon(\vec u) = \tfrac12( \nabla \vec u + \nabla \vec u ^T)$, $\vec C = \vec C(x,y,z)$ denotes the stiffness tensor, $\vec n$ means the unit normal vector and $\vec f$ stands for the pressure loads. $\Gamma_D$ is the Dirichlet and $\Gamma_N$ the Neumann boundary condition part of $\partial \Omega$. The stiffness tensor $\vec C$ describes the anisotropy of the material and represents its resistance against displacement under loads \cite{Gurtin73}. The PDE \eqref{pde} describes how an object made of linear materials changes its shape under given boundary conditions. Linear material means that the stress of the material depends linearly on the strain, and the stress is less than the yield strength of the material \cite{Slaughter12}. Deformation analysis is paramount for understanding the strain and stress distribution of the object, which can be used to predict structural failures.

The PDE \eqref{pde} requires numerical methods for its solution due to the absence of analytical solutions and the complexity of the domain ($\Omega$) \cite{numericalsolution}. In our study, $\Omega$ is the entire region the denture occupies. Because of the complex computational geometry and ellipticity of the PDE, we select FEM to approximate the displacement function through linear basis functions \cite{numericalsolution}. 

\subsection{Design geometry of the prosthesis and boundary conditions}
We use Ansys Inc.\ software as a finite element solver. The domain $\Omega$ is taken from a scanned three-dimensional facet model of the denture and transformed into a three-dimensional solid model along with the $\Gamma_D$ and $\Gamma_N$. The solid model is meshed to a three-dimensional unstructured mesh with a selected mesh size. These steps refer to the first two boxes in Figure \ref{fig:flowchart}. 

\subsection{Preparation of the reinforcement}
We use E-glass (strong material) as a reinforcement for our model. We optimize reinforcement distribution with topology optimization in the next step (fourth box in Figure \ref{fig:flowchart}). The Young's modulus and Poisson's ratio of the E-glass are given in Table \ref{materials}. These values are needed to construct the stiffness tensor $\vec C$ of the material to solve Equation \eqref{pde} \cite{Slaughter12}. We remark that unidirectional fiber reinforcement composites (unidirectional bundles of fibers) have anisotropic properties. However, we consider E-glass reinforcement an isotropic fiber to reinforce/strengthen the denture \cite{Zhang12}. This step refers to the third box in Figure \ref{fig:flowchart}. 

% 
%\begin{table}[!htb]\centering\small
 %\begin{tabular}{l|l|l}
  %\toprule
  %Material       &  PMMA  &   E-glass  \\
  %\midrule
  %Young´s modulus (MPa)   & 2550 Vallittu et al. \cite{Vallittu99}    & 72000 Safwat et al. \cite{SafwatKAK21} \\
  %Poisson´s ratio (--)   &  0.3 Ates et al. \cite{AtesCSSB06} & 0.2 Christiansson et al. \cite{christianssonH96}   \\
  %Density (g/cm³)  & 1.19 Vallittu et al. \cite{Vallittu99} & 2.54 Vallittu et al. %\cite{Vallittu99} \\
  %\bottomrule
 %\end{tabular}
 %\caption{Elastic properties of the materials we use in our model adopted from the literature.}\label{materials}
%\end{table}

\begin{table}[!htb]\centering
 \begin{tabular}{l|l|l}
  \toprule
  Material       &  PMMA  &   E-glass  \\
  \midrule
  Young's modulus (MPa)   & 2550 \cite{Vallittu99}    & 72000 \cite{SafwatKAK21} \\
  Poisson's ratio (--)   &  0.3 \cite{AtesCSSB06} & 0.2 \cite{christianssonH96}   \\
  Density (g/cm³)  & 1.19 \cite{Vallittu99} & 2.54 \cite{Vallittu99} \\
  \bottomrule
 \end{tabular}
 %\caption{Elastic properties of the materials we use in our model adopted from the literature.}
  \caption{Elastic properties of the materials used in our model. The values are adopted from the literature.}

 \label{materials}
\end{table}

%Next, we optimize the discretized PDE with ANSYS Topology Optimization 2021 R1's SIMP method, which is discussed in Section \ref{optimization}. During optimization, we remove a predefined portion of the strong material so the displacement remains minimal. Finally, we fill the removed strong material with the weak material parameters (which have lower mechanical properties), generating our final result. The whole procedure is summarized in Figure \ref{fig:flowchart}.

% We model the geometry with strong material and optimize it to find the optimal places of the reinforcement. After obtaining the places where reinforcement is needed, we add soft material in the void regions. To this end, we aim to decrease the average and maximum displacements within the model and determine the optimal reinforcement configuration in the domain to minimize compliance. The related flow chart can be found in Figure \ref{fig:flowchart}.\ARc{If one does not know what you do, one does not get an intuition from these sentences.}
\input{graph}

% \subsection{Numerical software}
% % 
% We design the denture's geometry in SpaceClaim 2021 R1 \cite{SpaceClaim21}. Afterward, we set the Dirichlet and Neumann boundary conditions in ANSYS Mechanical 2021 R1, we use the continuous Galerkin scheme in ANSYS Mechanical 2021 R1 for discretization. We optimize the structure using ANSYS Topology Optimization 2021 R1.

% \subsection{Modeling a baseline denture}
% % 
% We take a scanned three-dimensional facet model and transform it into a three-dimensional solid model with SpaceClaim 2021 R1. The geometry must be assigned Dirichlet, Neumann boundary conditions, and material. We assign the soft material for the baseline denture to investigate the displacement. Then, mesh size and element type must be specified in ANSYS Mechanical 2021 R1. Our focused output is to observe displacement without reinforcement.

\subsection{Topology optimization}
\sloppy
The rationale of the topology optimization algorithms is to find the optimal material distribution function $\theta(x,y,z): \Omega\to \{0,1\}$, ($\theta(x,y,z)=1$ indicates where the material is present) and voids ($\theta(x,y,z)=0$ indicates where the material is absent) under given constraints \cite{topologyandshape}. In this study, we use topology optimization to minimize the compliance of the reinforcement (strong material) while constraining the mass of the reinforcement. In other words, we seek the optimum of the following constrained optimization problem
\begin{subequations}\label{compliance}
	\begin{align}
	\min_{\theta} &\int_\Omega \vec \epsilon(\vec u(\theta)) : \vec C(\theta) : \vec \epsilon(\vec u(\theta)) \dx, \\
    \text{s.t}\;&\rho\int_\Omega \theta \dx  \le M_0, \label{compliance_constraint}
 \end{align}
\end{subequations}
where $\rho$ is the density of the strong material (reinforcement), and $M_0$ denotes the upper limit of the mass of the strong material. When the strong material distribution function $\theta=1$, the stiffness tensor $\vec C$ takes its normal value, corresponding to the presence of strong material. When $\theta=0$, the ersatz material approach is applied, meaning a virtual material with a very small Young's modulus is assigned to the stiffness tensor $\vec C$. This ersatz material simulates the absence of material while maintaining numerical stability to solve the linear elasticity PDE \cite{topologyandshape}. 
%Our objective is to find the material distribution $\theta(x,y,z)$ that minimizes the compliance of the prosthesis while constraining the mass of the reinforcement. 
As the topology optimization algorithm to solve \eqref{compliance}, we use the density-based SIMP method \cite{originalSIMP}. The SIMP method relaxes the strict binary requirement of $\theta$ having values of 0 or 1 for material distribution by allowing intermediate values. The intermediate values make this optimization task numerically more feasible by allowing the use of derivative-based optimization methods. Filtering techniques are used to regularize these intermediate values towards 0 or 1 \cite{originalSIMP}. SIMP method uses the sequential convex programming approach for solving the topology optimization problem under given constraints \cite{Zillober01}. We use the Ansys topology optimization module as a SIMP implementation. After topology optimization, we have a skeleton prosthesis, which consists of E-glass (reinforcement/strong material) and voids. This step refers to the fourth box in Figure \ref{fig:flowchart}.

\label{optimization}

\fussy
\subsection{Reinforced prosthesis}
The final step of the proposed method is to fill the voids of the skeleton prosthesis, which we get from the previous step, with the weak material (PMMA/resin base) to complete a reinforced prosthesis. In other words, we assign weak material (PMMA) to those regions in $\Omega$ where the optimized material distribution function is $\theta=0$. We emphasize that this step is not part of the topology optimization of the reinforcement which is done in the previous step. The Young's modulus and Poisson's ratio of the PMMA are given in Table \ref{materials}. Just like E-glass, we assume PMMA is isotropic \cite{Mounier12}. This step refers to the fifth box in Figure \ref{fig:flowchart}.

% We use the same boundary conditions, mesh size, and type for modeling the denture as in the baseline denture but we assign the strong material instead of the soft material. We use the SIMP method for optimization. Once the reinforcement is obtained, we reconstruct the optimized model by adding soft material to the region that SIMP removes. The contact types are assigned as bonded (connected). Boundary conditions are the same for each model since we aim to compare the baseline and reinforced dentures. Eventually, the model can be analyzed to obtain displacement with reinforcement. 

\section{Numerical examples of reinforcement of a denture}

We study the displacement of a denture with and without reinforcement. The reinforcement of the prosthesis is obtained through the proposed method in Section \ref{section2}. As a clarification, a non-reinforced denture is made of only PMMA, and a reinforced denture is made of PMMA and E-glass. A mesh convergence study is carried out to analyze how the mesh size impacts the accuracy of the FEM simulations for the denture.

\subsection{Finite element mesh and boundary conditions}
At first, we transform the scanned facet model of a denture (see Figure \ref{figure2}) into a three-dimensional solid model using SpaceClaim 2021 R1. We use a resolution of 0.5 mm for the facet of the prosthesis. The prosthesis is split into two bodies: the denture base and the teeth. Small faces are created on top of each tooth where the tooth forces can be applied. 

 We mesh the computational domain of the prosthesis with tetrahedral quadratic elements in ANSYS. Our model has 104168 nodes and 61108 elements (the bottom right subfigure in Figure \ref{figure2}). The denture base and the teeth are part of this conformal mesh. A Dirichlet boundary condition ($\vec u=\vec0$) is on the lower surface of the denture, where the denture typically contacts the mouth tissue (the top left subfigure in Figure \ref{figure2}). Forces (Neumann boundary conditions) are applied vertically on the teeth (the bottom left subfigure in Figure \ref{figure2}). The forces are 150\,N in the anterior area (front) \cite{150}, 450\,N in the premolar area (side) \cite{450}, and 900\,N in the posterior area (back) \cite{ozcaneffect, bite1}. 

%For comparison purposes, we apply the whole model with PMMA, a weak material.

%As the number of natural teeth decreases, the maximal bite force also decreases. For instance, removable partial denture wearers have a maximal bite force of 300~N, while complete denture wearers have forces of 180~N \cite{LassiHKK85, ferrariomaximal}.

%Chewing forces typically range from 50-80~N \cite{WaltimoK95}, while 
%Maximum bite forces have been studied extensively in literature \cite{WaltimoK94, WaltimoK95}. 
Earlier studies have reported high measured maximum biting forces \cite{WaltimoK95, WaltimoK94, bite1}. We choose such high chewing forces in our model to consider the maximum bite capacity of individuals. These high biting forces may appear in real-life scenarios, such as chewing tough foods or during stress conditions like intense chewing or clenching. This is essential for designing dentures that withstand high-load conditions, ensuring their stiffness. 
%because the maximum biting force is observed in the molar teeth, and the force decreases by a factor of 3-4 in the front teeth compared to the molar teeth \cite{bite}.

\subsection{Non-reinforced denture}
As a baseline, we evaluate the non-reinforced denture. 
The static structural analysis demonstrates that the maximum displacement of the non-reinforced denture is 0.192 mm, and its average displacement is 0.063 mm. The highest displacement magnitudes are mostly observed in the anterior and posterior areas (see Figure \ref{figure3}). Displacements of the magnitude of 0.1 mm are quite common in the non-reinforced denture in Figure \ref{fig:histogram}. Note that the displacement magnitudes exhibit a slight asymmetry due to anatomical differences between the right and left sides of the denture, which are not identical in Figure \ref{figure3}.

%The maximum unilateral biting force in the posterior region can reach 900~N, although chewing forces typically range from 50-80~N \cite{WaltimoK95}. As the number of natural teeth decreases, the maximal bite force also decreases. For instance, removable partial denture wearers have a maximal bite force of 300~N, while complete denture wearers have forces of 180~N \cite{LassiHKK85, ferrariomaximal}. 

%The forces used in this study are typical maximal biting forces measured unilaterally, which are commonly used in dental science literature, and this study is based on their values \cite{behrfracture,ozcaneffect}.

%Additionally, using high forces helps identify pain thresholds and physiological responses of dental structures, which is important for minimizing discomfort and damage in clinical applications \cite{WaltimoK95}. 
\subsection{Reinforced denture}

We apply the method described in Section \ref{section2} and exclude the teeth domain from the optimization region. In essence, we employ the optimization method only for the denture base. We use three different upper limits (Equation \eqref{compliance_constraint}) for the mass of the strong material (E-glass) in the topology optimization step. The limits are 20\%, 40\%, 57\%. The reinforcement shapes are in Figure \ref{figure4}, respectively. If the reinforcement has a 40\% mass constraint, the reinforcement is not fully needed in the anterior teeth (see top right in Figure \ref{figure4}). This is an interesting result from a dental perspective because all anterior teeth are subjected to a load. The reinforcement is simply connected if the mass limit is 57\% (horseshoe shape). This reinforcement distribution is shown on the bottom in Figure \ref{figure4}.

In the following, we investigate the displacement of the denture, which has 20\% reinforcement. The maximum displacement of the reinforced denture with 20\% reinforcement is 0.171 mm, and its average displacement is 0.035 mm (see Figure \ref{figure3}). There is approximately a 44\% decrease in the average displacement and an 11\% decrease in the maximum displacement. The reinforcement significantly reduces displacement magnitude in the posterior and premolar areas, as seen in Figure \ref{figure3}. Compared to the baseline denture without reinforcement, most of the displacement magnitudes are less than 0.05 mm in the reinforced denture, which can be seen in Figure \ref{fig:histogram}. In contrast to the non-reinforced dentures, the 0.1 mm displacement magnitude is rare in the reinforced denture.

%Next paragraph: Again, most of the first sentence repeats the discussion in Section 2. Same holds for the first sentence of paragraph 3.

%The strong material (E-glass) is assigned to the denture base and teeth to investigate reinforcement regions in the denture. All other properties such as mesh type and size, and boundary conditions are the same as in the previous section since we want to compare the displacement of the non-reinforced denture and reinforced denture. The tooth body is excluded from the optimization region, the denture base. 
%The optimization objective is to maximize global structure stiffness and observe the dependence on mass reduction of the distribution of material placement.

%80\%, 60\% and 43\% of mass are reduced from the denture base. 

%The reinforcement volume decreases from 9827.4 mm³ to 1835.6 mm³ once the program is set to reduce the mass by 80\%.

\newcommand{\customhspace}{}
    \begin{figure}[htb]
     \centering
     \includegraphics[width=\textwidth]{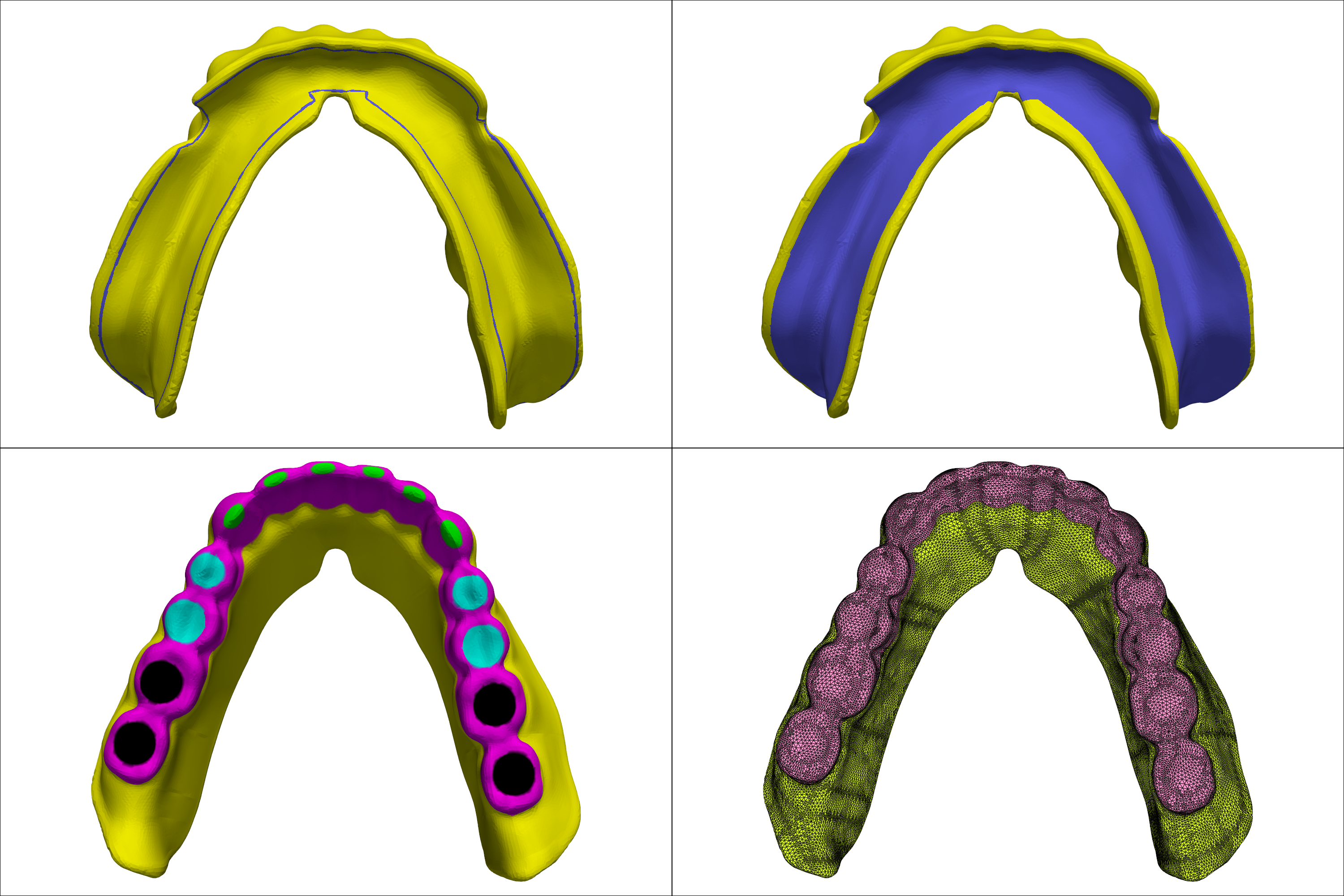}
       % \caption{}
       % \label{edgesupport}
    %\begin{subfigure}[b]{0.40\textwidth}
      %  \centering
     %   \includegraphics[width=\textwidth]{denture_support.png}
     %   \caption{}
   %     \label{edgesupport}
  %  \end{subfigure}
 %  \customhspace
  %   \begin{subfigure}[b]{0.40\textwidth}
  %       \centering
 %        \includegraphics[width=\textwidth]{facesupport.png}
 %        \caption{}
 %        \label{facesupport}
 %   \end{subfigure}
 %  \customhspace
%   \begin{subfigure}[b]{0.40\textwidth}
 %        \centering
%\includegraphics[width=\textwidth]{forces_denture.pdf}
    %     \caption{}
    %     \label{forces}
   % \end{subfigure}
   \customhspace
  
   \caption{A depiction of the boundary condition and forces of the denture. The figure on the top left is the fixed boundary of the denture (Dirichlet boundary) (blue line). The figure on the top right is the fixed boundary of the denture in the mesh convergence study (blue region). Forces are applied vertically (Neumann boundary) $900$\,N on posterior teeth (black region) $450$\,N on premolars (turquoise region) and $150$\,N on anterior teeth (green region) in the figure on the bottom left. The figure on the bottom right is the meshed denture.}
     \label{figure2}
      \end{figure}

      \begin{figure}
      \centering
      \includegraphics[width=\textwidth]{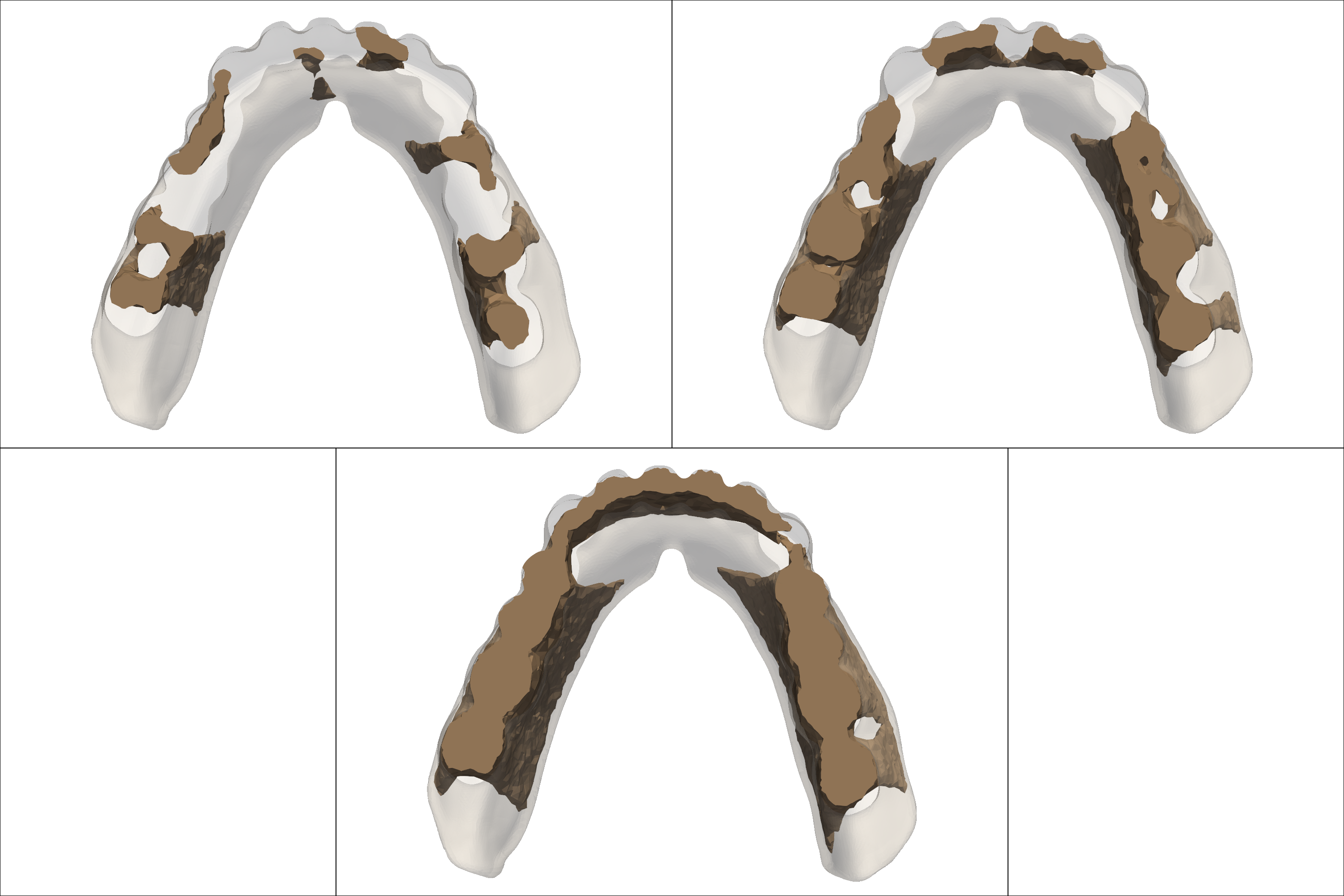}
    % \begin{subfigure}[b]{0.40\textwidth}
         
     %    \includegraphics[width=\textwidth]{TO20.png}
     %    \caption{}
     %    \label{80}
    % \end{subfigure}
   %\customhspace
   %  \begin{subfigure}[b]{0.40\textwidth}
   %      \centering
    %     \includegraphics[width=\textwidth]{TO40.png}
    %     \caption{}
    %     \label{60}
   %  \end{subfigure}
    %\customhspace
     %\begin{subfigure}[b]{0.40\textwidth}
      %   \centering
       %  \includegraphics[width=\textwidth]{TO-57.png}
        % \caption{}
         %\label{43}
     %\end{subfigure}
      \caption{%We present E-glass reinforcement (brown) distribution in this illustration. 
      An illustration of the E-glass reinforcement (brown) distribution for the optimized prostheses with varying upper limits of reinforcement mass. Left top figure: the reinforcement distribution with 20\% of the mass limit. Right top figure: the reinforcement distribution with 40\% of the mass limit. Bottom figure: the reinforcement distribution with 57\% of the mass limit. We use 20\% of the reinforcement in the denture displacement investigation.}
        \label{figure4}
     \end{figure}

         \begin{figure}[th!]
         \centering
    % \begin{subfigure}[b]{0.49\textwidth}
     %    \centering
      %   \includegraphics[width=\textwidth]{PMMA1.png}
     %    \caption{}
      %   \label{displacement-PMMA}
     % \end{subfigure}
    %\customhspace
    %\begin{subfigure}[b]{0.49\textwidth}
      %   \centering
     %    \includegraphics[width=\textwidth]{reinforcement1.png}
     %    \caption{}
     %    \label{displacement-e-glass}
    %  \end{subfigure}
      % \begin{subfigure}[b]{0.99\textwidth}
        % \centering
         \includegraphics[width=\textwidth]{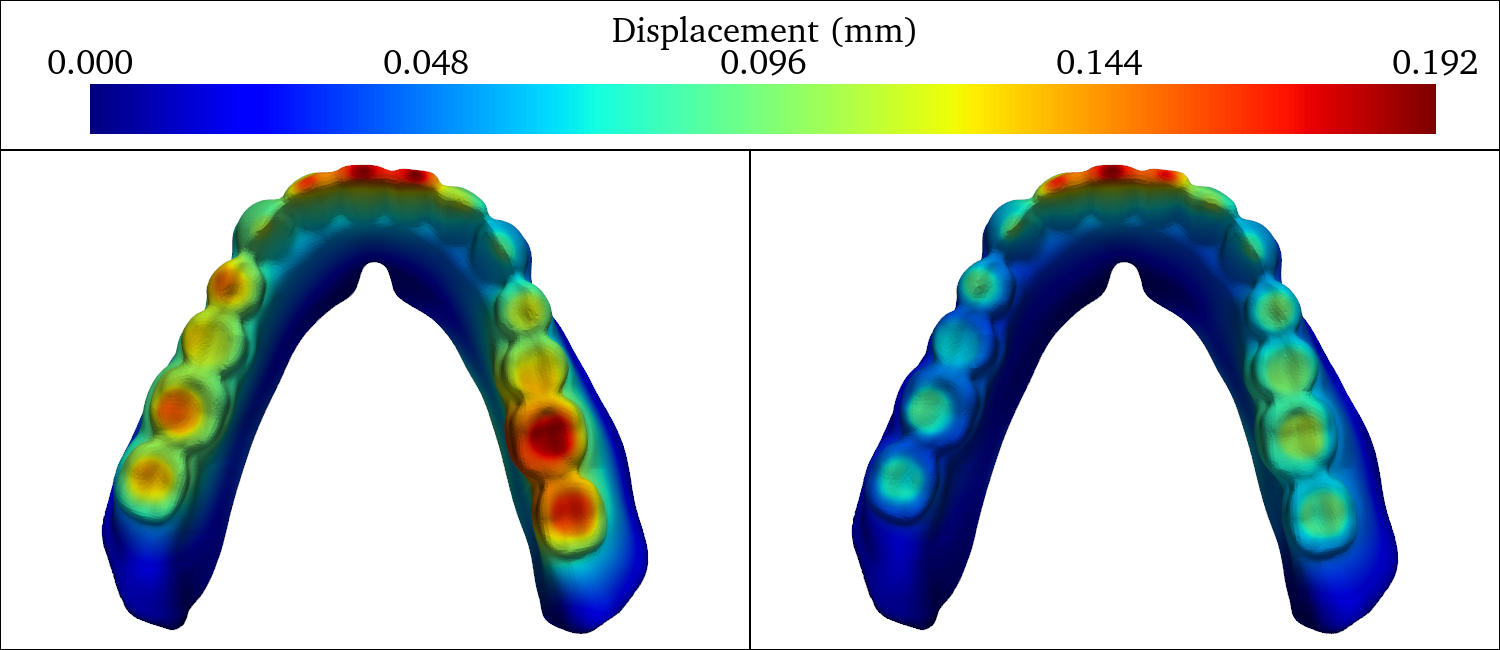}
        % \caption{}
        % \label{displacement-PMMA}
      %\end{subfigure}
      %\customhspace
      %    \begin{subfigure}[b]{0.49\textwidth}
         
        % \includegraphics[width=\textwidth]{mesh.png}
       %  \caption{}
        % \label{denture-mesh}
    %\end{subfigure}
   %\customhspace
      \caption{A visualization of the displacements of the dentures. The left-hand side image shows the displacement of the non-reinforced denture, and the right-hand side image illustrates the displacement of the reinforced denture. The maximum displacement of the non-reinforced denture is 0.192 mm. The maximum displacement of the reinforced denture is 0.17 mm, which is for the optimized 20\% of mass reinforcement. This comparison clearly shows the displacement of the non-reinforced (fully weak material) denture is higher than that of the reinforced denture.}

      %Some illustrations from simulations of the denture are demonstrated here. The left-hand side image visualizes the displacement of the non-reinforced denture, and the right-hand side image visualizes the displacement of the reinforced denture. The displacement result of the non-reinforced denture is 0.192 mm. The displacement result of the reinforced denture is 0.17 mm, which is for the optimized 20\% of mass reinforcement. 
      
      %The displacement of the denture exhibits a slight asymmetry due to anatomical differences between the right and left sides of the denture, which are not identical. This comparison clearly shows the displacement of the non-reinforced (fully weak material) denture is higher than that of the reinforced denture. In addition, we can point out that a denture has a higher mass percentage of optimized reinforcement like the top right illustration, and the down image in Figure \ref{figure4} can have smaller displacement (stiffer).

       \label{figure3}
      \end{figure}

\begin{figure}
    \centering
    \includegraphics[width=0.9\textwidth]{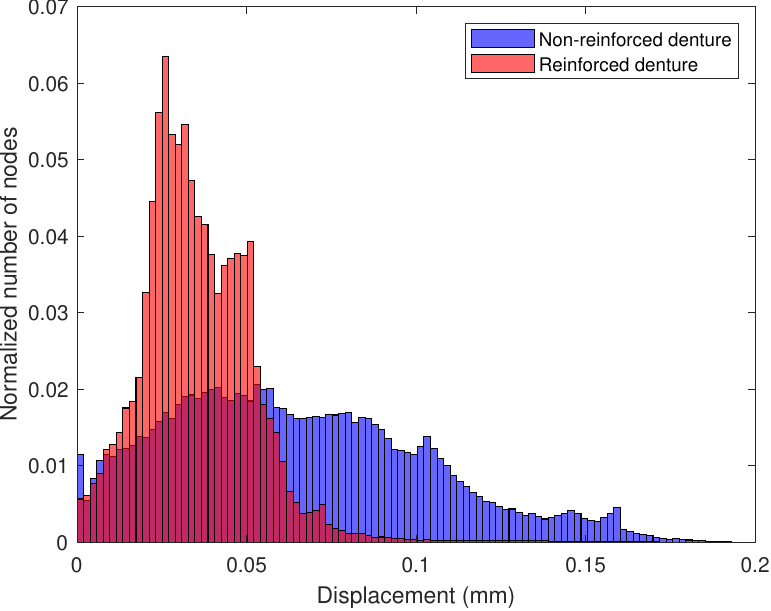}
    \caption{A presentation of the node-based displacement distributions of the non-reinforced and reinforced dentures. The reinforcement substantially reduces the high displacement magnitudes compared to the non-reinforced denture. This is evident in the node displacement distribution of the reinforced denture, which has a smaller variance and mean than the non-reinforced denture.}
    \label{fig:histogram}
\end{figure}
\FloatBarrier

\subsection{Mesh convergence study}

We investigate the effect of the mesh size on the accuracy of the simulation of a denture that consists of only E-glass. In other words, the denture does not contain PMMA, including the teeth body. A surface-fixed boundary is applied in the mesh convergence study (see Figure \ref{figure2} top right). An edge-fixed boundary is applied to the denture to minimize the effect of boundary conditions on the topology optimization results (see Figure \ref{figure2} top left). The forces are the same as before and given in Figure \ref{figure2}.

Our primal interest is the consistent total displacement. As seen in Table \ref{mesh-convergence}, the total displacement of the denture does not drastically change between the mesh sizes. We select the 0.5 mm mesh size for the non-reinforced and reinforced denture because the mesh size is computationally feasible in the topology optimization step. This mesh size also provides a good resolution with respect to the computational cost. Overall, the FEM results for the denture are reliable and independent of mesh size.

\begin{table}[!ht]
\centering
%\tiny
\small
\begin{tabular}{l|l|l|l|l}
   \toprule
 Mesh size  (mm)   &  2  &   1 &   0.5    & 0.25  \\
   \midrule
 %Maximum von-Mises (MPa) \cite{DuysinxS98,SvanbergW07}   &  42,634 & 42,04 & 42,126 & 41,621   \\
 
   Maximum principal stress (MPa)  &  20,781 & 20,639 & 24,364 & 34,8   \\
   Total displacement (mm)  & 0,00531 & 0,00533 & 0,0053  & 0,00541    \\
   Nodes & 16592 & 35396 & 112670 & 400940 \\
   
   Elements & 9153 & 20194 & 65903 & 239684 \\
   
    \bottomrule
\end{tabular}
\caption{\label{mesh-convergence}A mesh convergence study for the denture.}
\end{table}
% 

%The safety factor of 1 indicates a design's significant overcapacity. 
\section{Discussion}

%Average displacement is approximately 44\% decreasing, and maximum displacement is approximately 11\% decreasing.

%The displacement of dental prostheses in FEM is influenced by many factors, including the boundary conditions applied, the material properties of the prosthesis, and its geometry and shape. Dental restorative devices and prosthetic treatments must be designed to withstand forces similar to those experienced by natural teeth. 
\sloppy
In previous studies \cite{ShinyaY10,ShiF09}, FEM has been used to simulate the behavior of fiber-reinforced bridges to optimize their mechanical properties and stiffness. However, reinforcement of removable dentures has not previously been studied. Our motivation is to optimize the reinforcement of the denture robustly through a topology optimization framework. Thanks to this study, dental technicians can more productively fabricate the fiber reinforcements in the denture to minimize compliance. The study assists dental technicians in manufacturing reinforced dentures without resorting to time-consuming manual experimentation. The topology optimization method reduces material costs in manufacturing by using pieces of E-glass fiber reinforcement only where needed. %Additionally, this study allows for the fabrication of reinforced dentures using automated manufacturing techniques. 

Complete dentures are digitally fabricated using Computer-Aided Design (CAD) and Computer-Aided Manufacturing (CAM) technologies. These methods involve automated machinery for manufacturing, such as 5-axis milling machines or 3D printers, which could be used in conjunction with the proposed method in this study. However, the denture base materials employed in these digital techniques exhibit lower mechanical strength than traditional heat-cured acrylics used in conventional denture fabrication \cite{perea20213d,al2022printing}. Both digitally and traditionally manufactured prostheses can be reinforced with fibers to enhance their mechanical properties. Regardless of the fabrication method, it is essential to ensure that the reinforcing fibers are optimally positioned \cite{alander2021layer}. 
%\fussy
%narva2005static

%Bite forces in the posterior region can be as high as 900 newtons. They go down towards the anterior area. Forces used in this study are typical maximal biting forces measured unilaterally. However, during normal mastication, the forces are significantly lower, ranging from 50 to 80 N \cite{WaltimoK95, ferrariomaximal, ozcaneffect, behrfracture}. 

%Empirical design methods that rely on practical experience and observations instead of theoretical or computational models are commonly employed to optimize fiber reinforcement in dental prostheses. In traditional approaches, possible boundary conditions are evaluated for the denture, and the best result for fiber reinforcement is deemed optimal. The topology optimization method proposed in this work is a novel approach for structural design and placement of fiber reinforcements that utilizes computational algorithms to create optimized designs based on predetermined constraints and performance criteria together with first-line shape optimization methods.

  We provide a conceptual framework to optimize the reinforcement of the denture from the computational perspective, but practical manufacturing constraints might limit the feasibility of the optimized designs. This work is a continuation of our previous publication \cite{ECMS}, in which we presented a detailed description of the topology optimization-based reinforcement approach. In that paper, we demonstrated the method in the case of a two-dimensional numerical example with an emphasis on the finite element description of the optimization approach. We implemented the example in Julia and verified the numerical results using ANSYS Inc.\ software. 
 
 One motivation of the current study is to apply the methodology of our previous paper to a dental application that employs a three-dimensional unstructured mesh in the numerical model of the denture. Moreover, the denture is based on a real three-dimensional-scanned prosthesis sketch. Previous optimization studies of dental devices have often been designed by selecting the best option from a limited set of predefined designs \cite{ShiFQ08, ShinyaY10}. For example, the best option can refer to the design with the lowest compliance. Similarly, topology optimization has been used for dental implant applications to remove excessive material to make lightweight designs fabricated from only one material \cite{extension}. Our method extends the classical SIMP method to design multi-material dentures. In contrast to parametric approaches, our approach is non-parametric and naturally explores an extended range of potential reinforcement designs of the prosthesis without human assistance \cite{ShiF09}.

 Our study offers a foundation for future experimental studies. Because of the scope of the study, the denture reinforcement fabrication shall be addressed in a separate study. The manufacturability aspect could include the 3D printability of the reinforcements, including the problems related to the overhangs and clogging. We could also work on the printability of simpler geometries before considering the complex dentures. The shape distortions could be assessed and potentially tackled by CAD file precompensation. Printability issues might affect the surface finish of the dentures, which might result in drastically different final prostheses and their mechanical properties. Finally, the accuracy of the computed displacements of the prosthesis could be validated by comparing them to the displacement of the three-dimensional-printed multi-material reinforced dentures, whose composition was designed with our method.
 
 As a continuation of the current numerical study, we could also include stress in the optimization constraint. Stress constraints would be motivated by stress-induced damages, which can occur during high-impact force occasions. However, the current displacement study does not consider the stress of the reinforced dentures. Since stress investigation is numerically expensive, we focus on the displacement minimization of the denture as the objective and intentionally leave the stress constraints out of the scope. In future works, we could minimize the compliance under stress constraints. 
\fussy

\section{Conclusions}
\sloppy
We proposed a computational design method that optimizes the topology of the reinforcement for three-dimensional dentures with given mass constraints. One motivation of the study was to implement a reinforcement approach that identifies the optimal reinforcement material distribution for dentures. The proposed approach is general and eliminates the need for repetitive manual design processes. The method can deal with complex denture geometries, arbitrary boundary conditions, and two materials, which makes the method versatile and applicable. We optimized denture reinforcement to increase stiffness while minimizing material usage under mass constraints. In a numerical example, we minimized the displacement of a three-dimensional denture to enhance its stiffness. We performed a mesh sensitivity study to assess whether the numerical solution is accurate and independent of the finite element mesh size.

%We studied the reinforcement of the denture to minimize displacement and enhance stiffness. 

%By leveraging FEM, we can enable accurate simulation and analysis of mechanical behavior, and the predefined topology optimization allows the determination of the optimal reinforcement distribution. This combination of methodologies provides an efficient means of improving the design of dentures and enhancing their strength and durability. 

% (286) (asl) A new practical method was proposed computationally?

%(290)(asl) places? ----> region perhaps
\fussy

Future research should address how these optimized designs can be manufactured and ensure the practical applicability of reinforced dentures. Laboratory experiments could be performed to validate whether the displacements of the reinforced prosthesis are close to the predictions given by the numerical model. Exploring cost-effective production methods will be essential to make this technology accessible and personalized to patients. The removable partial dentures can also be studied since they have different constraints than removable complete dentures. Furthermore, we could minimize some other mechanical attributes of the prosthesis instead of its compliance. For instance, the weight of the prosthesis could be minimized under a constraint on the maximum allowed stress in the reinforced prosthesis to make it as light as possible while still being durable.

%The proposed method can also be applied to removable partial dentures.  

%Additionally, forces from the non-vertical axis can be applied to the denture in addition to the currently applied forces to investigate other possible load conditions.

%FUTURE IDEAS, OR SOME OTHER WAY RESULT DISCUSSION CRITICALLY, MANUFACTURA, VIVO (PERSONALIZED), NUMERICAL DISCRITIZATION.

%In future research, we will explore cost-effective manufacturing processes to make this technology accessible to patients.
 
\section*{Acknowledgements}
% 
%The Ansys licenses have been sponsored under the academic partnership program of EDRMedeso. We are grateful for the technical support and resources EDRMedeso and Ansys Inc.\ offer. 

The authors acknowledge EDRMedeso and Ansys Inc.\ for the software that contributed to this research, with licenses sponsored through the Academic Partnership program of EDRMedeso. The authors thank Dr.\ Mohamed Rabah for the technical support and Dr.\ Jarkko Suuronen for the useful discussion.

%The authors want to acknowledge EDRMedeso and Ansys Inc.\ for the software that has contributed to the success of this research. The licenses have been sponsored under the Academic Partnership program of EDRMedeso. The authors thank Dr.\ Mohamed Rabah for the technical support and Dr.\ Jarkko Suuronen for the useful discussion.
\section*{Funding}

%Full name: Lisäävän valmistuksen keskus bio- ja lääketeollisuudelle (AMBioPharma)
%Project code: A77805
%Funded by the European regional development fund (ERDF)
%This research was supported by funding from the AMBioPharma project, with project code A77805, and it was funded by the European Regional Development Fund (ERDF). The Research Council of Finland supported the work of R.\ Altunay and L.\ Roininen through the Flagship of Advanced Mathematics for Sensing, Imaging, and Modeling (decision number 359183); A.\ Rupp through Mathematical models and numerical methods for water management in soils (decision number 350101), Uncertainty quantification for PDEs on hypergraphs (decision number 354489), Localized orthogonal decomposition for high-order, hybrid finite elements (decision number 359633), and Finnish Flagship of advanced mathematics for Sensing, Imaging and Modeling, (decision number 358944). \textbf{(FIXME)} A.\ Rupp was also supported Business Finland's project number 539/31/2023 3D-Cure: 3D printing for personalized medicine and customized drug delivery.

This work has been funded by the European Regional Development Fund (project codes A77805, A80892), the Research Council of Finland (decision numbers 358944, 359183, 359633, 353095), and Business Finland (project numbers 539/31/2023, 147/31/2023).

\section*{Replication of results}
The files used in the simulation can be sent by request. 
\section*{Conflict of interest}
The authors state that there is no conflict of interest.
\section*{Ethical approval}
This article does not contain any studies with human participants or animals.
\section*{Declaration of generative AI and AI-assisted technologies in the writing process}
During the preparation of this work the authors used Grammarly in order to correct typographical errors and the grammar. After using Grammarly, the authors reviewed and edited the content as needed and took full responsibility for the content of the publication.

%\section*{Open Access}
 \bibliographystyle{elsarticle-num} 
 \bibliography{refs.bib}


%\begin{thebibliography}{00}


%\bibitem{}

%\end{thebibliography}
\end{document}

%% file: graph.tex
\begin{figure}[ht]
\centering 

\resizebox{\textwidth}{!}{%
\begin{tikzpicture}[
    % Define styles for the flowchart blocks, titles, and decision blocks
    outer_block/.style={draw, rectangle, fill=blue!15, minimum height=5cm, text width=\linewidth},
    outer_block_small/.style={draw, rectangle, fill=blue!15, minimum height=6cm, text width=4cm},
    outer_block_big/.style={draw, rectangle, fill=blue!15, minimum height=7cm, text width=10.5cm},
    title/.style={draw, rectangle, fill=blue, text=white, minimum height=0.5cm, align=center, font=\large},
    flowchart_block/.style={draw, rectangle, fill=white, text=blue, minimum height=3cm, text width=2cm, align=center, node distance=0.5cm},
    decision_block/.style={draw, diamond, fill=white, text=blue, minimum height=1cm, text width=2cm, align=center, node distance=1cm},
]

% First flowchart
\node[outer_block] (outer_block1) {}; % Create the first outer block
\node[title, anchor=north] at (outer_block1.north) (title1) {Proposed method for the reinforcement of the model}; % Add a title to the first outer block and define its position to the outer block

% Add flowchart blocks inside the first outer block
% Start by defining the first flowchart block to the following:
 anchor -> starting position
% xshift -> optional. Only used if you want to shift in the x direction
% below -> define its position relative to the title. More options can be used such as right, left, below right, etc.
\node[flowchart_block, anchor=north west, xshift=0.1cm, below=0.4cm of title1.south west] (block1) {Design geometry};
\node[flowchart_block, right=of block1] (block2) {Dirichlet and Neumann boundary conditions};
\node[flowchart_block, right=of block2] (block3) {Assign strong material};
\node[flowchart_block, right=of block3] (block4) {Topology optimization};
\node[flowchart_block, right=of block4] (block5) {Fill weak material to produce original shape};

% Draw arrows between flowchart blocks in the first flowchart
\draw[->] (block1) -- (block2);
\draw[->] (block2) -- (block3);
\draw[->] (block3) -- (block4);
\draw[->] (block4) -- (block5);
\end{tikzpicture}
}
\caption{Steps of the proposed reinforcement method for the dental prosthesis.}
\label{fig:flowchart}
\end{figure}

%% file: main.bbl
\begin{thebibliography}{10}
\expandafter\ifx\csname url\endcsname\relax
  \def\url#1{\texttt{#1}}\fi
\expandafter\ifx\csname urlprefix\endcsname\relax\def\urlprefix{URL }\fi
\expandafter\ifx\csname href\endcsname\relax
  \def\href#1#2{#2} \def\path#1{#1}\fi

\bibitem{imeko1}
I.~Papallo, M.~Martorelli, F.~Lamonaca, A.~Gloria, Generative design and
  insights in strategies for the development of innovative products with
  tailored mechanical and/or functional properties, Acta IMEKO 12~(4) (12
  2023).

\bibitem{lecturenote1}
I.~Papallo, A.~Gloria, M.~Martorelli, Design of additive manufactured devices
  with tailored properties: Tackling biomedical challenges, in: Design Tools
  and Methods in Industrial Engineering III, Springer Nature Switzerland, 2024,
  pp. 77--83.

\bibitem{compositereview}
H.~Mehboob, S.-H. Chang, Application of composites to orthopedic prostheses for
  effective bone healing: A review, Composite Structures 118 (2014) 328--341.

\bibitem{imeko2}
P.~Fucile, F.~Lamonaca, A.~Gloria, L.~Moroni, A methodological approach towards
  the bio-inspired design of novel scaffolds for tissue engineering, Acta IMEKO
  12~(4) (12 2023).

\bibitem{BioTriz}
J.~F. Vincent, O.~A. Bogatyreva, N.~R. Bogatyrev, A.~Bowyer, A.-K. Pahl,
  Biomimetics: its practice and theory, Journal of the Royal Society Interface
  3~(9) (2006) 471--482.

\bibitem{toporeview}
J.~Zhu, H.~Zhou, C.~Wang, L.~Zhou, S.~Yuan, W.~Zhang, A review of topology
  optimization for additive manufacturing: Status and challenges, Chinese
  Journal of Aeronautics 34~(1) (2021) 91--110.

\bibitem{structuraldesign}
A.~A. Al-Tamimi, H.~Almeida, P.~Bartolo, Structural optimisation for medical
  implants through additive manufacturing, Progress in Additive Manufacturing
  5~(2) (2020) 95--110.

\bibitem{sridhar2010optimal}
I.~Sridhar, P.~Adie, D.~Ghista, Optimal design of customised hip prosthesis
  using fiber reinforced polymer composites, Materials \& Design 31~(6) (2010)
  2767--2775.

\bibitem{chethan2019finite}
K.~Chethan, M.~Zuber, S.~Shenoy, et~al., Finite element analysis of different
  hip implant designs along with femur under static loading conditions, Journal
  of Biomedical Physics \& Engineering 9~(5) (2019) 507.

\bibitem{topologyandshape}
G.~Allaire, C.~Dapogny, F.~Jouve, Shape and topology optimization, in:
  {Geometric partial differential equations, part II}, Vol.~22 of Handbook of
  Numerical Analysis, Elsevier, 2021, pp. 1--132.

\bibitem{lecturenotes2}
P.~Ausiello, M.~Martorelli, I.~Papallo, A.~Gloria, R.~Montanari, M.~Richetta,
  A.~Lanzotti, Optimal design of surface functionally graded dental implants
  with improved properties, in: Advances on Mechanics, Design Engineering and
  Manufacturing IV, Springer International Publishing, 2023, pp. 294--305.

\bibitem{sun2021additive}
B.~Sun, Q.~Ma, X.~Wang, J.~Liu, M.~Rejab, Additive manufacturing in medical
  applications: a brief review, in: IOP Conference Series: Materials Science
  and Engineering, Vol. 1078, IOP Publishing, 2021, p. 012007.

\bibitem{additive}
C.~Comotti, D.~Regazzoni, C.~Rizzi, A.~Vitali, Additive manufacturing to
  advance functional design: an application in the medical field, Journal of
  Computing and Information Science in Engineering 17~(3) (2017) 031006.

\bibitem{lecturenotes3}
C.~de~Crescenzo, M.~Richetta, I.~Papallo, P.~Fucile, M.~Martorelli, A.~Gloria,
  A.~Lanzotti, Surface roughness prediction in fused deposition modeling: An
  engineered model, in: Design Tools and Methods in Industrial Engineering III,
  Springer Nature Switzerland, 2024, pp. 101--108.

\bibitem{materials}
R.~B. Osman, M.~V. Swain, A critical review of dental implant materials with an
  emphasis on titanium versus zirconia, Materials 8~(3) (2015) 932--958.

\bibitem{NarvaVHY01}
K.~K. Narva, P.~K. Vallittu, H.~Helenius, A.~Yli-Urpo, Clinical survey of
  acrylic resin removable denture repairs with glass-fiber reinforcement,
  International Journal of Prosthodontics 14~(3) (2001).

\bibitem{LadizeskyCW90}
N.~Ladizesky, T.~Chow, I.~Ward, The effect of highly drawn polyethylene fibres
  on the mechanical properties of denture ease resins, Clinical Materials 6~(3)
  (1990) 209--225.

\bibitem{VallittuL92reinforcement}
P.~Vallittu, V.~Lassila, Reinforcement of acrylic resin denture base material
  with metal or fiber strengtheners, Journal of Oral Rehabilitation 19~(3)
  (1992) 225--230.

\bibitem{fracture}
D.~Smith, The acrylic denture mechanical evaluation mid-line fracture, Br.
  Dent. J. 110 (1961) 257--267.

\bibitem{reinforcement}
N.~Ladizesky, C.~Ho, T.~Chow, Reinforcement of complete denture bases with
  continuous high performance polyethylene fibers, The Journal of Prosthetic
  Dentistry 68~(6) (1992) 934--939.

\bibitem{perea20213d}
L.~Perea-Lowery, M.~Gibreel, P.~K. Vallittu, L.~V. Lassila, 3d-printed vs.
  heat-polymerizing and autopolymerizing denture base acrylic resins, Materials
  14~(19) (2021) 5781.

\bibitem{Zafar20}
M.~S. Zafar, Prosthodontic applications of polymethylmethacrylate {(PMMA)}: An
  update, Polymers 12~(10) (2020) 2299.

\bibitem{JomjunyongRRACK17}
K.~Jomjunyong, P.~Rungsiyakull, C.~Rungsiyakull, W.~Aunmeungtong,
  M.~Chantaramungkorn, P.~Khongkhunthian, Stress distribution of various
  designs of prostheses on short implants or standard implants in posterior
  maxilla: a three-dimensional finite element analysis, Oral \& Implantology
  10~(4) (2017) 369.

\bibitem{reinforcement1}
D.~Jagger, A.~Harrison, K.~Jandt, The reinforcement of dentures, Journal of
  Oral Rehabilitation 26~(3) (1999) 185--194.

\bibitem{ShinyaY10}
A.~Shinya, D.~Yokoyama, Finite Element Analysis for Dental Prosthetic Design,
  INTECH Open Access Publisher, Sciyo, 2010.

\bibitem{ShiF09}
L.~Shi, A.~Fok, Structural optimization of the fiber-reinforced composite
  substructure in a three-unit dental bridge, Dental Materials 25~(6) (2009)
  791–801.

\bibitem{ShiFQ08}
L.~Shi, A.~Fok, A.~Qualtrough, A two-stage shape optimization process for
  cavity preparation, Dental Materials 24~(11) (2008) 1444--1453.

\bibitem{originalSIMP}
M.~P. Bend{\o}e, O.~Sigmund, Topology optimization: theory, methods, and
  applications, Springer Science \& Business Media, 2003.

\bibitem{extension}
C.-L. Chang, C.-S. Chen, C.-H. Huang, M.-L. Hsu, Finite element analysis of the
  dental implant using a topology optimization method, Medical Engineering \&
  Physics 34~(7) (2012) 999--1008.

\bibitem{ECMS}
R.~Altunay, J.~Suuronen, A.~Rupp, E.~Immonen, L.~Roininen, Reinforcement
  approach using topology optimization, in: ECMS, 2024, pp. 330--337.

\bibitem{hvejsel2011material}
C.~F. Hvejsel, E.~Lund, Material interpolation schemes for unified topology and
  multi-material optimization, Structural and Multidisciplinary Optimization 43
  (2011) 811--825.

\bibitem{Gurtin73}
M.~E. Gurtin, The linear theory of elasticity, Linear Theories of Elasticity
  and Thermoelasticity: Linear and Nonlinear Theories of Rods, Plates, and
  Shells (1973) 1--295.

\bibitem{Slaughter12}
W.~S. Slaughter, The linearized theory of elasticity, Springer Science \&
  Business Media, 2012.

\bibitem{numericalsolution}
P.~G. Ciarlet, The Finite Element Method for Elliptic Problems, Society for
  Industrial and Applied Mathematics, 2002.

\bibitem{Zhang12}
M.~Zhang, J.~P. Matinlinna, E-glass fiber reinforced composites in dental
  applications, Silicon 4 (2012) 73--78.

\bibitem{Vallittu99}
P.~K. Vallittu, Flexural properties of acrylic resin polymers reinforced with
  unidirectional and woven glass fibers, The Journal of Prosthetic Dentistry
  81~(3) (1999) 318--326.

\bibitem{SafwatKAK21}
E.~M. Safwat, A.~G. Khater, A.~G. Abd-Elsatar, G.~A. Khater, Glass
  fiber-reinforced composites in dentistry, Bulletin of the National Research
  Centre 45 (2021) 1--9.

\bibitem{AtesCSSB06}
M.~Ate{\c{s}}, A.~Cilingir, T.~S{\"u}l{\"u}n, E.~S{\"u}nb{\"u}lo{\u{g}}lu,
  E.~Bozda{\u{g}}, The effect of occlusal contact localization on the stress
  distribution in complete maxillary denture, Journal of Oral Rehabilitation
  33~(7) (2006) 509--513.

\bibitem{christianssonH96}
H.~Christiansson, J.~Helsing, Poisson’s ratio of fiber-reinforced composites,
  Journal of Applied Physics 79~(10) (1996) 7582--7585.

\bibitem{Zillober01}
C.~Zillober, Global convergence of a nonlinear programming method using convex
  approximations, Numerical Algorithms 27 (2001) 265--289.

\bibitem{Mounier12}
D.~Mounier, C.~Poil{\^a}ne, C.~B{\^u}cher, P.~Picart, Evaluation of transverse
  elastic properties of fibers used in composite materials by laser resonant
  ultrasound spectroscopy, in: Acoustics 2012, 2012, pp. 1246--1250.

\bibitem{150}
E.~Hellsing, C.~Hagberg, Changes in maximum bite force related to extension of
  the head, The European Journal of Orthodontics 12~(2) (1990) 148--153.

\bibitem{450}
Z.~D. Soliman, S.~S. Bedair, Comparing the fracture resistance of three
  different cad/cam occlusal veneers manufactured with different thicknesses.
  an in vitro study., Egyptian Dental Journal 70~(2) (2024) 1813--1826.

\bibitem{ozcaneffect}
M.~Ozcan, M.~H. Breuklander, P.~K. Vallittu, The effect of box preparation on
  the strength of glass fiber-reinforced composite inlay-retained fixed partial
  dentures, The Journal of Prosthetic Dentistry 93~(4) (2005) 337--345.

\bibitem{bite1}
P.~d.~S. Calderon, E.~M. Kogawa, J.~R.~P. Lauris, P.~C.~R. Conti, The influence
  of gender and bruxism on the human maximum bite force, Journal of Applied
  Oral Science 14 (2006) 448--453.

\bibitem{WaltimoK95}
A.~Waltimo, M.~K{\"o}n{\"o}nen, Maximal bite force and its association with
  signs and symptoms of craniomandibular disorders in young {F}innish
  non-patients, Acta Odontologica Scandinavica 53~(4) (1995) 254--258.

\bibitem{WaltimoK94}
A.~Waltimo, M.~K{\"o}n{\"o}nen, Bite force on single as opposed to all
  maxillary front teeth, European Journal of Oral Sciences 102~(6) (1994)
  372--375.

\bibitem{al2022printing}
F.~D. Al~Qarni, M.~M. Gad, Printing accuracy and flexural properties of
  different 3d-printed denture base resins, Materials 15~(7) (2022) 2410.

\bibitem{alander2021layer}
P.~Alander, L.~Perea-Lowery, K.~Vesterinen, A.~Suominen, E.~S{\"a}ilynoja,
  P.~K. Vallittu, Layer structure and load-bearing properties of fibre
  reinforced composite beam used in cantilever fixed dental prostheses, Dental
  Materials Journal 40~(1) (2021) 165--172.

\end{thebibliography}
